# Carrier Transport in Films of Alkyl-Ligand-Terminated Silicon Nanocrystals

Ting Chen,[1] Brian Skinner,[2,3] Wei Xie,[1] B. I. Shklovskii**,[2] and Uwe R. Kortshagen*,[4]

[1]Department of Chemical Engineering and Materials Science, University of Minnesota, 421 Washington Ave SE, Minneapolis, Minnesota 55455, United States

[2]Fine Theoretical Physics Institute, University of Minnesota, Minneapolis, MN 55455, United States

[3]Materials Science Division, Argonne National Laboratory, Argonne, IL 60439, United States

[4]Department of Mechanical Engineering, University of Minnesota, 111 Church Street SE, Minneapolis, Minnesota 55455, United States

ABSTRACT: Silicon nanocrystals (Si NCs) have shown great promise for electroluminescent and photoluminescent applications. In order to optimize the properties of Si NC devices, however, electronic transport in Si NCs films needs to be thoroughly understood. Here we present a systematic study of the temperature and electric field dependence of conductivity in films of alkyl-ligand-terminated Si NCs, which to date have shown the highest potential for device applications. Our measurements suggest that the conductivity is limited by the ionization of rare NCs containing donor impurities. At low bias, this ionization is thermally activated, with an ionization energy equal to twice the NC charging energy. As the bias is increased, the ionization energy is reduced by the electric field, as determined by the Poole-Frenkel effect. At large bias and sufficiently low temperature, we observe cold ionization of electrons from donor-containing NCs, with a characteristic tunneling length of about 1 nm. The temperature- and electric-field-dependent conductance measurements presented here provide a systematic and comprehensive picture for electron transport in lightly doped nanocrystal films.

**INTRODUCTION**

Semiconductor nanocrystals (NCs) have great potential for thin-film optoelectronics, such as solar cells[1] and light emitting diodes,[2] due to their size-tunable electronic properties[3,4] and solution processability.[5] Among the different materials for semiconductor NCs, silicon (Si) has attracted substantial interest because of its abundance and low toxicity. Significant progress has been made in developing synthetic methods to prepare high quality Si NCs,[6] achieving controllable doping,[7–9] and integrating NCs into high performance optoelectronic devices. Such devices include hybrid organic-Si NC solar cells[10] and light emitting devices with external quantum efficiency approaching 9%.[11] Since most device applications rely on electrical conduction through films of NCs, understanding the fundamental mechanisms of the carrier transport in NC films is necessary to improve device performance.

While a number of studies have examined electrical conduction in group II-VI semiconductor NCs,[12–14] the electronic transport in Si NC films is still poorly understood. Most transport studies thus far have focused on Si NCs embedded in an oxide matrix or covered by an oxide shell, and in these systems Fowler-Nordheim tunneling,[15] space-charge-limited current (SCLC),[15,16] and hopping[17] transport mechanisms have been identified. Electrical conduction in H-terminated Si NCs has also been studied, and SCLC[18] and hopping[19] conduction mechanisms have been considered to describe the experimental results. However, for alkyl-ligand-terminated Si NCs, which to date have exhibited the best performance in photoluminescent and electroluminescent applications,[10,20] the fundamental electronic transport mechanisms have not yet been studied. A fundamental study of the electronic transport mechanisms in this system is therefore needed.

In this work, we investigate the electronic transport in thin films of alkyl-ligand-terminated Si NCs by studying the temperature- and electric-field-dependent conductance. A vertical two-terminal structure is employed, and the electrical transport in the NC films is examined over a range of temperatures (300 ~ 10 K). While these studies have been performed with a focus on films of Si NCs, we expect that many of our conclusions apply to other nanocrystal materials as well.

**EXPERIMENTAL METHODS**

The Si NCs for which data are shown were synthesized in a nonthermal rf (13.56 MHz) plasma reactor, as described previously.[21] The reactor was a Pyrex tube with 10-mm outer diameter which then expands to 25-mm outer diameter, and the reactor pressure was held constant during synthesis at 1.4 Torr. The flow rates of precursors were 13 sccm (standard cubic centimeters per minute) of silane and 35 sccm of argon.

Hydrogen gas was injected at 100 sccm in the 25 mm expansion region of the reactor tube. Particle size was controlled by adjusting the argon flow rate. The nanoparticles have spherical shape and are virtually 100 % crystalline. To achieve stable colloids, the Si NC surfaces were functionalized with alkyl ligands in a liquid-phase thermal hydrosilylation reaction with a 5:1 mixture of mesitylene and 1-dodecene.[22] A clear colloidal dispersion of particles was obtained after a 2 h refluxing period at 215 ˚C. After functionalization, the alkyl ligands were covalently bonded to the particle surface and individually dispersed Si NCs were produced.[23] Before device fabrication, the alkylated Si NCs were purified by dispersion/precipitation for three rounds. It is well known that the introduction of a nonsolvent that is miscible with the original dispersing solvent can destabilize the NC dispersion and cause NCs to aggregate and precipitate.[5] Therefore, the synthetic by-products or unreacted ligands are left in the solution, helping to produce a clean NC material. For alkylated Si NCs, anhydrous chloroform and acetonitrile were used as the solvent and nonsolvent, respectively, and a centrifugation process at 4600 rpm for 15 min was performed after each precipitation. After purification the Si NCs were redispersible in nonpolar solvents (e.g., chloroform). All solvents used in the synthesis and functionalization process were well dried and degassed to avoid oxidation of NCs.

The crystallinity and the size of Si NCs were characterized by X-ray diffraction (XRD) using a Bruker-AXS microdiffractometer with a 2.2-kW sealed Cu X-ray source at 40 kV and 40 mA (wavelength 0.154 nm). The XRD pattern was recorded for drop-cast films of 1-dodecene alkylated Si NCs on a glass substrate. The high resolution bright field transmission electron microscopy (HRTEM) employed FEI Tecnai G2 F-30 TEM with a Schottky field-emission electron gun operated at 100 kV accelerating voltage. The sample for TEM was prepared by drop casting a submonolayer of Si NCs directly onto a copper lacey carbon TEM grid covered with a continuous carbon film (5 nm thick). The morphology of the spin-coated Si NC films was studied by a Bruker NanoScope V Multimode scanning probe microscope working in tapping mode. The probes used were silicon cantilevers with integrated tapping mode tips fabricated by NanoWorld AG (Arrow$^{TM}$ NCR, resonant frequency 285 kHZ, spring constant 42 N/m). Cross-sectional scanning electron microscopy (SEM) images were obtained by a field emission gun JEOL 6500 with accelerating voltage 5 kV and working distance 10 mm.

Vertical two-terminal devices were constructed on microscope glass slides in a nitrogen-filled glovebox. The glass substrates were pre-cleaned by sequential ultrasonication for 10 min each in acetone, methanol and isopropyl alcohol, and were dried by nitrogen blow. The glass was then placed in a M. Braun thermal evaporator and 30 nm thick Al strips were deposited through a shadow mask on the glass to serve as the bottom electrodes. Following the deposition of the electrodes, Si NCs films were spin-coated from a 50 mg/ml dispersion in anhydrous chloroform at 1500 rpm for 60 s. The resulting film thickness was about 180 nm, determined from the cross-sectional SEM images. Finally, Al top electrodes with 150 nm thickness were thermally evaporated through another shadow mask on top of the Si NC films with 2 mm$^2$ and 4 mm$^2$ active areas, thus finishing the devices. The deposition rates used for both electrodes were controlled at 0.5 Å/s. The devices were then transferred into another nitrogen-filled glovebox for subsequent electrical measurements without any air exposure.

The current-voltage (*I-V*) characteristics of the NC films were recorded in a Desert Cryogenics (Lakeshore) probe station in a nitrogen-filled glovebox with Keithley 236 and 237 source measuring units and homemade LabVIEW programs. Low temperature measurements employed a Lakeshore 331 temperature controller with a fixed ramp rate of 2 K/min. All measurements were carried out in the dark and under vacuum at pressure ~ $10^{-3}$ Torr.

## RESULTS AND DISCUSSION

**A. Structural Characterization of Si NCs.** The crystallinity and particle size of the alkylated (hydrosilylated) Si NCs were examined using X-ray diffraction (XRD) and a high resolution transmission electron microscope (HRTEM). Figure 1a shows a typical wide-angle out-of-plane *2θ-ω* diffraction pattern (coupled scan) for drop-cast 1-dodecene alkylated Si NCs. Well-defined peaks indicating silicon diamond structure were observed, and the mean NC diameter was estimated to be ~ 4 nm using the Scherrer equation.[24] The NC size was also confirmed from high resolution TEM image as shown in Figure 1b, and the standard deviation of the size distribution is within ~ 10% of the peak size. Clear lattice fringes indicate that particles are single crystalline.

**B. Time and Temperature Dependence of Conductance.** NC films spin-coated onto glass substrates pre-patterned with bottom Al electrodes were smooth, continuous, and devoid of pinholes, as shown in the atomic force microscopy (AFM) image Figure 2a. Electrical measurements were conducted with the two-point probe geometry as shown in Figure 2b. A cross-sectional SEM image of the device is displayed in Figure 2c, in which the electrodes and Si NC film are labeled. The thickness of the Si NC film is about 180 nm. At low bias, the current-voltage (*I-V*) characteristics show ohmic behavior, so that the conductance $G = I/V$ is voltage-independent. The presence of a measurable ohmic response suggests that, even though our films are not intentionally doped, there is a noticeable concentration of carriers in the system independent of the drive voltage. This indicates the presence of a certain level of unintentional doping in the film. This picture is confirmed by measurements of the temperature dependence of the conductance, presented below.

The measured conductance of our films at room temperature is plotted in Figure 3a as a function of time. During these measurements, the film was kept inside the nitrogen-filled glovebox with oxygen and water levels less than 0.1 ppm and *I-V* curves were recorded at different times *t* (with $t = 0$ corresponding to freshly made films). The NC film conductance exhibits a clear "aging" effect, with *G* increasing by more than two orders of magnitude over the course of the first 2 week and then remaining almost constant over several months. Freshly made films show conductance that varies from $10^{-9}$ to $10^{-8}$ S and the conductance of completely-aged films reaches $10^{-7}$ to $10^{-6}$ S. Figure 3b displays the current-voltage characteristics of films measured as freshly made and after aging for two months; ohmic behavior can be seen in both curves. We attribute the aging effect to an increase in the tunneling rate between NCs, likely caused by a slow physical rearrangement of NCs or adsorption of water molecules, as we discuss in more detail below.

The temperature dependence of the ohmic conductance was measured for each film when it was freshly made, and after storing inside the glovebox for 1 week and for 2 week, since these time periods represented three typical stages of aging. For the following discussion, we refer to these three stages as freshly made, 1-week-aged and 2-week-aged. The ohmic conductance was measured with a fixed low voltage bias (~ 100 mV) while cycling the temperature from 300 to 120 K. For each sample, we find that the temperature-dependent conductance data is reversible: the conductance returns to its original values after warming up to 300 K, indicating that the sample remains stable during the temperature cycling.

In general, the conductance of an insulating disordered system, like our NC films, decreases with decreasing temperature and can typically be described by

$$G \propto \exp\left[-\left(\frac{T_0}{T}\right)^\gamma\right], \quad (1)$$

where the temperature exponent $\gamma$ depends on the transport mechanism and $T_0$ is a characteristic temperature.[25] Situations with $\gamma = 1$, in general, correspond to nearest-neighbor hopping (NNH), where the conductance arises primarily from tunneling events between neighboring NCs. On the other hand, $\gamma = 0.5$ and $\gamma = 0.25$ correspond to Efros-Shklovskii variable-range hopping (ES-VRH) and Mott variable-range hopping (M-VRH), respectively. These mechanisms arise in situations where the current is carried primarily by "cotunneling" events between non-neighboring NCs. In different temperature regimes, different charge transport mechanisms can dominate.[25]

*I-V* curves in the low bias regime for the 2-week-aged film are displayed in Figure 4a for a range of different temperatures. No hysteresis is observed between the forward and reverse voltage sweeps, and *I-V* curves are highly symmetric around the origin, implying that the ohmic behavior persists for all measured temperatures. Similar behavior is observed in the films that were freshly made and 1-week-aged. For all films, the conductance decreases with decreasing temperature, and we observe a linear dependence of ln $G$ on $T^{-1}$, suggesting $\gamma = 1$, for all three aging stages. The data can, in principle, be plotted against either $T^{-1}$ or $T^{-1/2}$ in order to compare which of these dependencies gives a more consistent fit. In our case, however, these plots alone are not enough to make clear the power of $T$ because the temperature range is limited by the very small current, as seen in Figure 4b. Here we interpret our data with an Arrhenius dependence. As we show below, this interpretation is consistent with the electric field dependence of the conductivity.

Figure 4b displays Arrhenius plots of the conductance versus temperature for the same Si NC film measured at each of the three stages. As discussed above, the observation of $\gamma = 1$ in these plots suggests that the carrier transport corresponds to NNH throughout the measured interval of temperature. In this case, the temperature $T_0$ in eq 1 can be associated with an activation energy $E_a = T_0/k$, where $k$ is Boltzmann's constant, so that

$$G \propto \exp\left[-\frac{E_a}{kT}\right]. \quad (2)$$

The value of the activation energy is obtained by measuring the slope of the Arrhenius plot of conductance. This process gives $E_a = 135 \pm 13$ meV, $180 \pm 2$ meV and $165 \pm 2$ meV for the freshly made, 1-week-aged and 2-week-aged stages, respectively. These activation energies are much smaller than the Si band gap, which indicates that the current is carried by a finite, temperature-independent concentration of electrons within the conduction band rather than by electrons activated from the valence band. In other words, our films have a noticeable amount of unintentional doping, as mentioned above. The source of this doping remains unclear, although, as we discuss below, it is apparently at the level of less than about one donor (or acceptor) per NC. Below we will assume that we deal with donors.

The activation energy of ~ 160 meV that appears in our films is associated with the energy required to ionize a donor-containing NC. To understand the activation energy that arises from this ionization process, consider first that in the ground state of the system every conduction band electron resides in one of the rare NCs that contain a positive donor charge, and so every NC is electro-neutral. When such a donor-containing NC is ionized and its conduction band electron is removed to a distant NC that does not contain a donor (an "empty" NC), the electron becomes free from donors and is able to hop freely between nearest-neighboring NCs. Such free electrons are responsible for the film's conductivity. In this ionized state the system has two charged NCs, one with a positive donor and another with a negative electron, and so the ionization process costs an energy $2E_c$, where $E_c$ is the "charging energy" associated with adding a net charge $e$ to an electrically neutral NC. In equilibrium, such ionization processes occur with the same rate as electron-donor recombination processes. The latter rate is proportional to $n^2$, where $n$ is the concentration of free electrons on empty NCs. As a result of this equilibration the activation energy associated with the concentration $n$ and the conductance $G$ is equal to $E_c$.

By way of analogy, we note that if one compares the NC array described above with a lightly doped n-type bulk semiconductor, then empty NCs play the role of the conduction band, in the sense that they provide the states through which free electrons carry the system current. Donor-containing NCs play the role of conventional donors, and the ionization energy $2E_c$ is analogous to the donor binding energy. The concentration of electrons on empty NCs, $n$, plays the role of the electron concentration in the conduction band.

The value of $E_c$ can be estimated by noting that when a charge is introduced to the interior of an NC, it causes dielectric polarization of the interior of the NC and the surrounding dielectric environment. Since the internal dielectric constant $\varepsilon_{NC}$ is much larger than the external dielectric constant $\varepsilon_i$, the great majority of a given internal charge is distributed at the surface of the NC by the dielectric response. The resulting Coulomb self-energy of the NC can therefore be thought of as equivalent to that of a metallic sphere in a uniform dielectric medium.[26] This allows one to estimate the charging energy as

$$E_c = \frac{e^2}{4\pi\varepsilon_0\varepsilon_r D} , \quad (3)$$

where $D$ is the particle diameter, $\varepsilon_0$ is the permittivity of vacuum and $\varepsilon_r$ is the effective dielectric constant for the NC film. It has been pointed out that $\varepsilon_r$ of the NC array is not simply the dielectric constant $\varepsilon_i$ of the insulating medium between NCs, but that the effect of neighboring NCs' polarization response to an applied field should also be considered. In this case, the effective dielectric constant of the NC film can be estimated from the canonical Maxwell-Garnett formula[27]

$$\varepsilon_r \approx \varepsilon_i \frac{\varepsilon_{NC} + 2\varepsilon_i + 2f(\varepsilon_{NC} - \varepsilon_i)}{\varepsilon_{NC} + 2\varepsilon_i - f(\varepsilon_{NC} - \varepsilon_i)}, \quad (4)$$

where $f$ is the volume fraction of NCs and $\varepsilon_{NC} = 11.7$ is the dielectric constant of Si. Since the dielectric constant of organic ligands is much lower than that of Si and these ligands fill a relatively small fraction of the system volume, we can safely neglect the dielectric response of these ligands and set $\varepsilon_i = 1$. From the refractive index measurement, the film density $f$ was determined as 42%, and this gives $\varepsilon_r = 2.46$. If one assumes that NCs within our film are arranged in a random close packing, then the spacing $\delta$ between the surfaces of neighboring particles is about 0.6 nm. This separation between particles is reasonable because the 1.4 nm-long organic ligands on the particle surface should largely lie flat against the surface in order to achieve lower energy states. Inserting $\varepsilon_r = 2.46$ into eq 3 gives an estimate for the charging energy of $E_c = 146$ meV, which is in good agreement with the experimentally measured value of the activation energy.

Our observation of NNH conduction for all films, regardless of age, suggests that for all of these films the level of (unintentional) doping is less than about one donor per NC. Indeed, in the opposite case of large doping, NCs become charged in the ground state to avoid electron occupation of energetically expensive 1P quantum states of the NC,[26] and the activated behavior is replaced by ES-VRH with temperature exponent $\gamma = 1/2$. Our films show no sign of this $\gamma = 1/2$ behavior, and so we conclude that the level of doping is small.

We now turn our attention to the aging effect of the conductance. As shown above, the activation energies for the films as freshly made and 2-week-aged are close, while the conductance at these two stages varies by two orders of magnitude. In general, such an increase in the conductance could arise either from an increase in the number of carriers or from a reduction of the tunneling barrier between neighboring NCs, which enhances the characteristic frequency of electron tunneling events. An increase in the donor concentration with film age seems unlikely, since chemical changes in the NCs can occur only at the surface, and surface states generally create deep mid-gap states rather than shallow donor levels. There is also no evidence that our films are heavily compensated by any such deep surface states, since this compensation would lead to a pinning of the chemical potential to the donor levels and therefore to an activation energy $2E_c$ rather than $E_c$.[25] We therefore conclude that the increase in conductivity comes from a reduction in the tunneling gaps between neighboring NCs. Such a reduction could arise, for example, if the spacing between NCs were gradually reduced as leftover interstitial solvent molecules evaporate and NCs readjust. The tunneling barrier could also be reduced without rearrangement of NCs, for example by adsorption of water molecules in the interstitial gaps, which would provide easier tunneling pathways for electrons. The effects of adsorbed water on film conductivity for semiconductors and oxides have been reported previously,[28] and Rastgar *et al.* observed conductivity increase by up to an order of magnitude due to water adsorption in thin films of Si NCs.[19]

As mentioned above, the origin of the free carriers in our films is not clear yet. The incorporation of impurities during synthesis is one possible source, but further work is needed to confirm this hypothesis.

The aging effect may also be accelerated by the weak exposure to sunlight, which is known to produce charge separation of carriers. A detailed exploration of the effect of light exposure on device aging is left for a later work.

**C. Electric Field Dependence of Conductance.** As discussed above, thermally activated NNH provides a good description of our data at low bias voltage, within the regime of ohmic response. Here we present results for conductivity at larger bias, focusing our attention on films that have been aged for 2 week. At a temperature of 300 K, the *I-V* curve for bias voltage between 0 and 20 V shows significant hysteresis, as displayed in Figure 5a. The forward scan is also plotted in log-log scale in Figure 5b. This hysteretic behavior is generally observed in systems where injected charge carriers become trapped. In our measurements this hysteresis was dramatically suppressed at lower temperatures (below 260 K), which suggests that carrier trapping is thermally activated. The nearly complete disappearance of hysteresis at low temperatures indicates that the electron transport is not affected by traps at these low temperatures, and that below 260 K the electron transport mainly occurs through the interior quantum states of each particle rather than through particle surface trap states.[13]

The temperature dependence of the film *I-V* characteristics at high bias is shown in Figure 6a for temperatures ranging from 260 to 80 K. Below 80 K, the *I-V* curves are temperature independent within the measurable voltage range and coincide with the curve corresponding to 80 K. To interpret our measurements, we first consider the temperature range of 260 to 80 K. In the low field regime (below 1 V), the slope of the *I-V* curves in log-log scale is close to 1, which indicates ohmic behavior. As the temperature is decreased, the current in the ohmic regime decreases according to the Arrhenius law, as described in the previous section. For high electric fields (bias voltage above 1 V), the current acquires a stronger dependence on voltage until about 16 V, when the *I-V* curves for different temperatures converge. In the regime between 1 and 16 V, the *I-V* characteristics are strongly non-ohmic, but retain an Arrhenius-like dependence on temperature. Figure 6b displays the conductance *G* versus *T* at various voltages from 300 to 10 K. Activation energies in the temperature range 300 to 80 K at each voltage point can be extracted from the slope of each curve, and are plotted in Figure 6c. With increasing voltage, the activation energy is seen to decrease. This behavior is likely a manifestation of the Poole-Frenkel effect,[29,30] which describes the reduction of the ionization energy for a donor-containing NC by application of an electric field *F* (see Figure 7a and b). In particular, such an electric field creates an addition to the Coulomb potential of the donor that depends linearly on distance, so that the electric potential has a maximum in the direction opposite the electric field at a distance $r_{max} = (e/4\pi\varepsilon_0\varepsilon_r F)^{1/2}$. The value of this potential maximum determines the energy necessary to ionize the donor, which becomes $2E_c - 2(e^3 F / 4\pi\varepsilon_0\varepsilon_r)^{1/2}$ as shown in Figure 7b. Consequently, at large enough voltage that the reduction in the ionization energy is much larger than *kT*, the conductance follows[30]

$$G \propto \frac{kT}{\sqrt{e^3 F / 4\pi\varepsilon_0\varepsilon_r}} \exp\left[\frac{-E_c + \sqrt{e^3 F / 4\pi\varepsilon_0\varepsilon_r}}{kT}\right], (5)$$

where, again, the activation energy is equal to half the donor ionization energy. This reduction in the apparent activation energy by an amount proportional to the square root of the applied voltage indeed provides a good description of our data, as shown in Figure 6c. We note that at sufficiently large fields, $r_{max}$ becomes smaller than the spacing $D+\delta$ between neighboring NCs, and eq 5 should lose its validity. At such large fields the ionization energy is determined by the energy required to activate the electron to the nearest-neighboring NC in the direction opposite the field as shown in Figure 7c, and the activation energy decreases linearly with the applied voltage as shown by the dash line in Figure 6c. This large-voltage regime can be called the Poole limit of Poole-Frenkel effect and corresponds to $V$ larger than ~ 5 V in our samples.

At even larger bias voltages, $V$ greater than ~ 16V, the ionization energy is completely overwhelmed by the "downstream" electric potential, and the activation energy for the conductance is eliminated, so that the current becomes essentially temperature-independent. As one can see, at such large fields the conductance at $T = 80$ K is enhanced by more than 6 orders of magnitude relative to its value at 1V.

We now turn our attention to the low temperature regime, $T < 80$ K, where the observed current is essentially independent of temperature at all bias voltages for which the current is measurable. At such small temperatures the thermal energy is insufficient to ionize donor-containing NCs. Instead, cold ionization processes dominate the conductance. In particular, at small enough biases that the electric field $F$ is much smaller than a critical field $F_0 = \dfrac{2E_c}{e(D+\delta)}$, the electron tunnels directly from donor-containing NC to an empty NC at a distance $x \sim 2E_c/eF$.[31] This tunneling distance depends on the applied electric field, with higher electric field implying shorter tunneling distance, as illustrated in Figure 7d (For example, at a bias voltage of 4 V, the tunneling distance $x \sim$ 14.4 nm, which is about three NC spacings.) The resulting ionization rate is proportional to $\exp\left(-\frac{2x}{\xi}\right)$, where $x$ is the hopping distance and $\xi$ is the characteristic tunneling length (localization length). As a result, the conductance is

$$G \propto \exp\left[-\frac{x}{\xi}\right]. \quad (6)$$

Rewriting eq 6 with $x \sim 2E_c/eF$ gives

$$G \propto \exp\left[-\frac{2E_c}{e\xi F}\right]. \quad (7)$$

Therefore, the characteristic tunneling length $\xi$ can be extracted from a linear fit of $\ln G$ versus $1/F$. In Figure 6d, $\ln G$ is shown to depend linearly on $1/F$ in the range of $5 < V < 16$ V, and the corresponding characteristic length $\xi$ is determined to be $1.1 \pm 0.1$ nm. This characteristic tunneling length describes the typical distance over which the electron wave function is localized spatially, and is related to the decay length $a$ of the electron wave function in the gap between two NCs. In particular, when an electron tunnels to a distant, non-neighboring NC at a distance $x$, its tunneling trajectory involves passing through a chain of intermediate NCs (see Figure 7c), and the decay of the electron tunneling amplitude is dominated by passage through the gaps between neighboring NCs along the chain. As a consequence, the tunneling amplitude is suppressed by a factor $\sim \exp\left[\left(-\frac{2\delta}{a}\right) \times \left(\frac{x}{D+\delta}\right)\right]$, so that the characteristic tunneling length is given by $\xi \sim a\left(\frac{D+\delta}{\delta}\right)$.[32] From this relation we estimate that the decay length $a$ for the hydrosilylated SiNC film is $\sim 0.14$ nm, which is similar in magnitude to previous measurements of the tunneling decay length in PbSe NCs[33] and simple estimates using the work function of Si.

At lower electric field (V < 5 V), the film conductance $G$ deviates from the linear relation of ln $G$ versus $1/F$ at large electric field, which could be caused by the presence of rare, atypical NC pairs with very close spacing. Such "bridges" may play a role in our geometry because we drive current across a thin film between large area contacts.[34]

As is the case for larger temperatures, at sufficiently large electric fields that $F > F_0$, which corresponds to $V \geq 16$ V, the bias voltage is strong enough to generate electrons by inducing tunneling directly between neighboring NCs.

Even though our explanation of the experimental data has used the language of donors and conduction electrons in our films, the analysis applies to acceptors and conducting holes as well. Unfortunately, due to the extremely low mobility of carriers in Si NCs,[35] it is probably not easy to verify the type of dopants from Hall measurements or transistor measurements. To further investigate this unintentional doping in our materials, one could introduce a small amount of intentional n-type or p-type dopants into the Si NCs and deduce the carrier type for the unintentionally doped sample from the change in the film conductivity.

**CONCLUSIONS**

In summary, we have investigated the temperature and electric field dependent electron transport in thin films of alkyl-ligand-terminated Si NCs. Our interpretation of the data is based on assumption that a small fraction of NCs contains a donor. At low bias, the film conductance is determined by electrons activated from donor-containing NCs to empty NCs, between which they move via nearest neighbor hopping. As the bias is increased, the conductance is enhanced by the electric field, which causes a reduction in the activation energy as determined by Poole-Frenkel effect. At large bias and sufficiently low temperature, we observe cold ionization of donor-containing NCs via electron tunneling to distant empty NCs with characteristic tunneling length 1 nm. Compared with other Si NCs used in electrical measurements, the light doping and good surface passivation of our alkyl ligands-terminated Si NCs provide a good platform to study carrier transport in a lightly doped nanocrystal film, which, to the best of our knowledge, has not been explored previously. The unintentional doping that appears in our Si NCs enables us to probe the electronic transport under low electric field, and the good surface passivation ensures that the measurements were not hindered by surface defects. Our work constitutes the first thorough study of carrier transport in a lightly doped nanocrystal film, and we believe that many of the results obtained here for Si NCs apply to other NC materials as well.

**ACKNOWLEDGEMENTS**

We are grateful to K. V. Reich for valuable comments. This work was supported primarily by the National Science Foundation through the University of Minnesota MRSEC under Award Number DMR-0819885. Part of this work was carried out in the College of Science and Engineering Characterization Facility, University of Minnesota, which has received capital equipment funding from the NSF through the UMN MRSEC program. Work at Argonne National Laboratory was supported by the U.S. Department of Energy, Office of Basic Energy Sciences under contract no. DE-AC02-06CH11357.

**AUTHOR INFORMATION**

**Corresponding Author**

*E-mail: Kortshagen@umn.edu, phone (612)-626-2542

**E-mail: shklovsk@physics.umn.edu, phone (612)-625-0771

**REFERENCES**

(1) Kamat, P. V. Quantum Dot Solar Cells. Semiconductor Nanocrystals as Light Harvesters†. *J. Phys. Chem. C* **2008**, *112*, 18737–18753.
(2) Wood, V.; Bulović, V. Colloidal Quantum Dot Light-Emitting Devices. *Nano Rev.* **2010**, *1*, 1–7.
(3) Klimov, V. I. *Semiconductor and Metal Nanocrystals: Synthesis and Electronic and Optical Properties*; CRC Press, 2003.
(4) Alivisatos, A. P. Semiconductor Clusters, Nanocrystals, and Quantum Dots. *Science* **1996**, *271*, 933–937.
(5) Murray, C. B.; Kagan, C. R.; Bawendi, M. G. Synthesis and Characterization of Monodisperse Nanocrystals and Close-Packed Nanocrystal Assemblies. *Annu. Rev. Mater. Sci.* **2000**, *30*, 545–610.
(6) Mangolini, L.; Thimsen, E.; Kortshagen, U. High-Yield Plasma Synthesis of Luminescent Silicon Nanocrystals. *Nano Lett.* **2005**, *5*, 655–659.
(7) Stegner, A. R.; Pereira, R. N.; Lechner, R.; Klein, K.; Wiggers, H.; Stutzmann, M.; Brandt, M. S. Doping Efficiency in Freestanding Silicon Nanocrystals from the Gas Phase:

Phosphorus Incorporation and Defect-Induced Compensation. *Phys. Rev. B* **2009**, *80*, 165326.

(8) Pi, X. D.; Gresback, R.; Liptak, R. W.; Campbell, S. A.; Kortshagen, U. Doping Efficiency, Dopant Location, and Oxidation of Si Nanocrystals. *Appl. Phys. Lett.* **2008**, *92*, 123102.

(9) Rowe, D. J.; Jeong, J. S.; Mkhoyan, K. A.; Kortshagen, U. R. Phosphorus-Doped Silicon Nanocrystals Exhibiting Mid-Infrared Localized Surface Plasmon Resonance. *Nano Lett.* **2013**, *13*, 1317–1322.

(10) Cheng, K.-Y.; Anthony, R.; Kortshagen, U. R.; Holmes, R. J. High-Efficiency Silicon Nanocrystal Light-Emitting Devices. *Nano Lett.* **2011**, *11*, 1952–1956.

(11) Liu, C.-Y.; Holman, Z. C.; Kortshagen, U. R. Hybrid Solar Cells from P3HT and Silicon Nanocrystals. *Nano Lett.* **2009**, *9*, 449–452.

(12) Guyot-Sionnest, P. Electrical Transport in Colloidal Quantum Dot Films. *J. Phys. Chem. Lett.* **2012**, *3*, 1169–1175.

(13) Kang, M. S.; Sahu, A.; Norris, D. J.; Frisbie, C. D. Size- and Temperature-Dependent Charge Transport in PbSe Nanocrystal Thin Films. *Nano Lett.* **2011**, *11*, 3887–3892.

(14) Yu, D.; Wang, C.; Wehrenberg, B. L.; Guyot-Sionnest, P. Variable Range Hopping Conduction in Semiconductor Nanocrystal Solids. *Phys. Rev. Lett.* **2004**, *92*, 216802.

(15) Burr, T. A.; Seraphin, A. A.; Werwa, E.; Kolenbrander, K. D. Carrier Transport in Thin Films of Silicon Nanoparticles. *Phys. Rev. B* **1997**, *56*, 4818–4824.

(16) Rafiq, M. A.; Tsuchiya, Y.; Mizuta, H.; Oda, S.; Uno, S.; Durrani, Z. A. K.; Milne, W. I. Charge Injection and Trapping in Silicon Nanocrystals. *Appl. Phys. Lett.* **2005**, *87*, 182101.

(17) Rafiq, M. A.; Tsuchiya, Y.; Mizuta, H.; Oda, S.; Uno, S.; Durrani, Z. A. K.; Milne, W. I. Hopping Conduction in Size-Controlled Si Nanocrystals. *J. Appl. Phys.* **2006**, *100*, 014303.

(18) Pereira, R. N.; Niesar, S.; You, W. B.; da Cunha, A. F.; Erhard, N.; Stegner, A. R.; Wiggers, H.; Willinger, M.-G.; Stutzmann, M.; Brandt, M. S. Solution-Processed Networks of Silicon Nanocrystals: The Role of Internanocrystal Medium on Semiconducting Behavior. *J. Phys. Chem. C* **2011**, *115*, 20120–20127.

(19) Rastgar, N.; Rowe, D. J.; Anthony, R. J.; Merritt, B. A.; Kortshagen, U. R.; Aydil, E. S. Effects of Water Adsorption and Surface Oxidation on the Electrical Conductivity of Silicon Nanocrystal Films. *J. Phys. Chem. C* **2013**, *117*, 4211–4218.

(20) Jurbergs, D.; Rogojina, E.; Mangolini, L.; Kortshagen, U. Silicon Nanocrystals with Ensemble Quantum Yields Exceeding 60%. *Appl. Phys. Lett.* **2006**, *88*, 233116.

(21) Anthony, R. J.; Rowe, D. J.; Stein, M.; Yang, J.; Kortshagen, U. Routes to Achieving High Quantum Yield Luminescence from Gas-Phase-Produced Silicon Nanocrystals. *Adv. Funct. Mater.* **2011**, *21*, 4042–4046.

(22) Mangolini, L.; Jurbergs, D.; Rogojina, E.; Kortshagen, U. Plasma Synthesis and Liquid-Phase Surface Passivation of Brightly Luminescent Si Nanocrystals. *J. Lumin.* **2006**, *121*, 327–334.

(23) Mangolini, L.; Jurbergs, D.; Rogojina, E.; Kortshagen, U. High Efficiency Photoluminescence from Silicon Nanocrystals Prepared by Plasma Synthesis and Organic Surface Passivation. *Phys. Status Solidi C* **2006**, *3*, 3975–3978.

(24) Scherrer, P. Estimation of the Size and Internal Structure of Colloidal Particles by Means of Röntgen. *Nachr Ges Wiss Gött.* **1918**, *2*, 96–100.

(25) Efros, A. L.; Shklovskii, B. I. *Electronic Properties of Doped Semiconductors*; Springer-Verlag, New York, 1984.

(26) Skinner, B.; Chen, T.; Shklovskii, B. I. Theory of Hopping Conduction in Arrays of Doped Semiconductor Nanocrystals. *Phys. Rev. B* **2012**, *85*, 205316.

(27) Maxwell, J. C. *A Treatise on Electricity and Magnetism*; Clarendon press, 1881; Vol. 1.

(28) Traversa, E. Ceramic Sensors for Humidity Detection: The State-of-the-Art and Future Developments. *Sens. Actuators B Chem.* **1995**, *23*, 135–156.

(29) Frenkel, J. On Pre-Breakdown Phenomena in Insulators and Electronic Semi-Conductors. *Phys. Rev.* **1938**, *54*, 647.
(30) Hartke, J. L. The Three-Dimensional Poole-Frenkel Effect. *J. Appl. Phys.* **1968**, *39*, 4871–4873.
(31) Shklovskii, B. I. Hopping Conduction in Semiconductors Subjected to a Strong Electric Field. *Sov Phys Semicond.* **1973**, *6*, 1964.
(32) Zhang, J.; Shklovskii, B. I. Density of States and Conductivity of a Granular Metal or an Array of Quantum Dots. *Phys. Rev. B* **2004**, *70*, 115317.
(33) Liu, Y.; Gibbs, M.; Puthussery, J.; Gaik, S.; Ihly, R.; Hillhouse, H. W.; Law, M. Dependence of Carrier Mobility on Nanocrystal Size and Ligand Length in PbSe Nanocrystal Solids. *Nano Lett.* **2010**, *10*, 1960–1969.
(34) Levin, E. I.; Ruzin, I. M.; Shklovskii, B. I. Transverse Hopping Conductivity of Amorphous Films in Strong Electric Fields. *Sov Phys Semicond.* **1988**, *22*, 401–408.
(35) Gresback, R.; Kramer, N. J.; Ding, Y.; Chen, T.; Kortshagen, U. R.; Nozaki, T. Controlled Doping of Silicon Nanocrystals Investigated by Solution-Processed Field Effect Transistors. *ACS Nano* **2014**, *8*, 5650–5656.

**FIGURES**

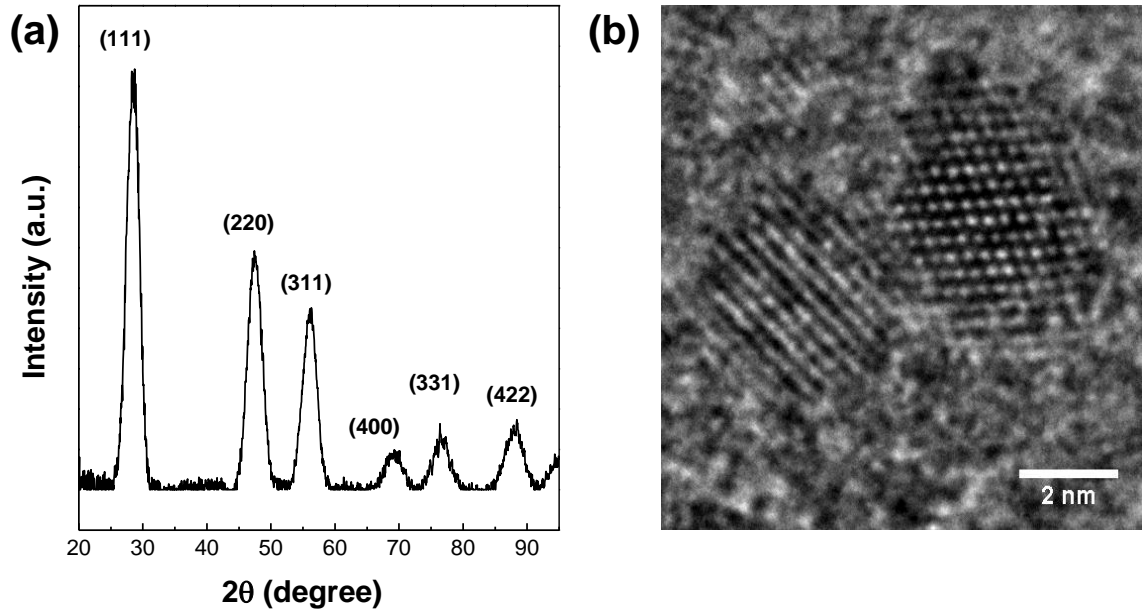

Figure 1. (a) XRD pattern and (b) High Resolution TEM image for Si NCs alkylated with 1-dodecene in a solution mixture of 5:1 mesitylene:1-dodecene.

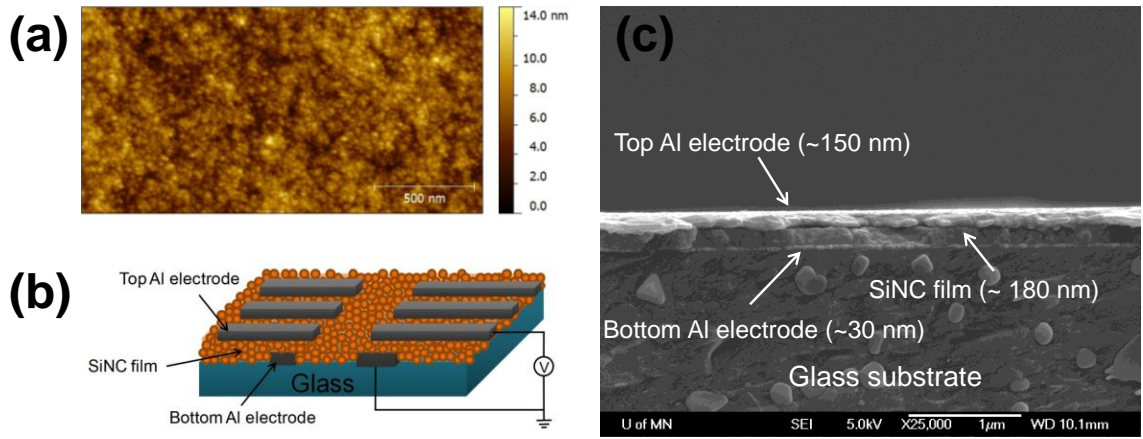

Figure 2. (a) AFM height image of 1-dodecene alkylated Si NC films. (b) Schematic cross-sectional diagram of the two-terminal vertical structure of our devices (not to scale). The active areas are square 2 mm$^2$ and 4 mm$^2$, respectively. (c) SEM image of the device cross section.

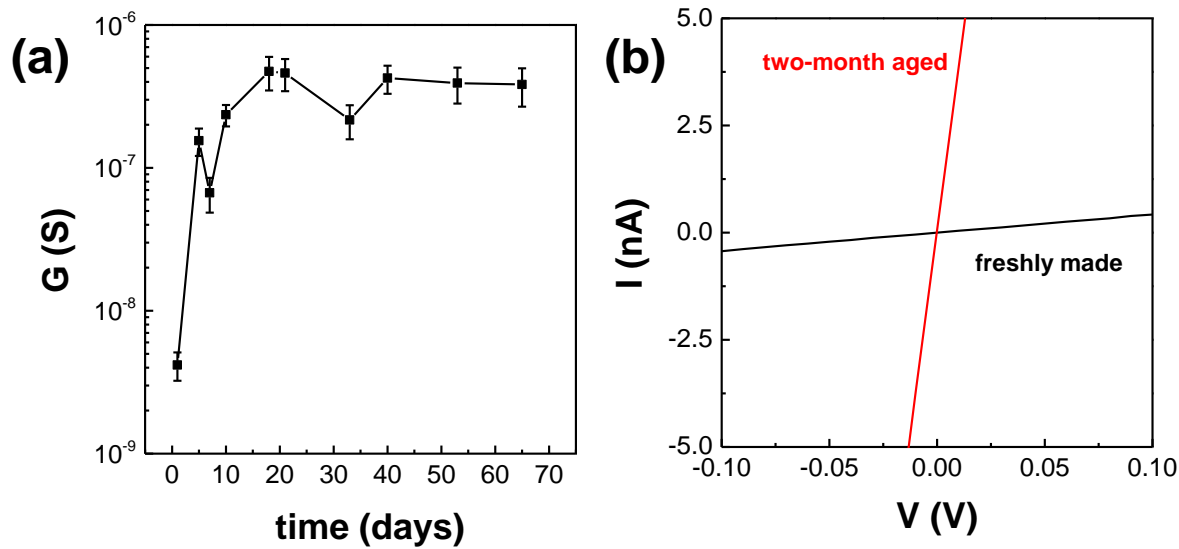

Figure 3. (a) Time dependence of electrical conductance for thin films of 1-dodecene alkylated Si NCs at room temperature as the films were kept inside a nitrogen-filled glovebox (oxygen and water level below 0.1 ppm). Black squares represent the average conductance of five devices at each time point, with error bars indicating the standard deviation. (b) Current-Voltage (*I-V*) characteristics measured for a typical Si NC film as freshly made ($t = 0$) and after aging at room temperature after two months ($t = 65$ days).

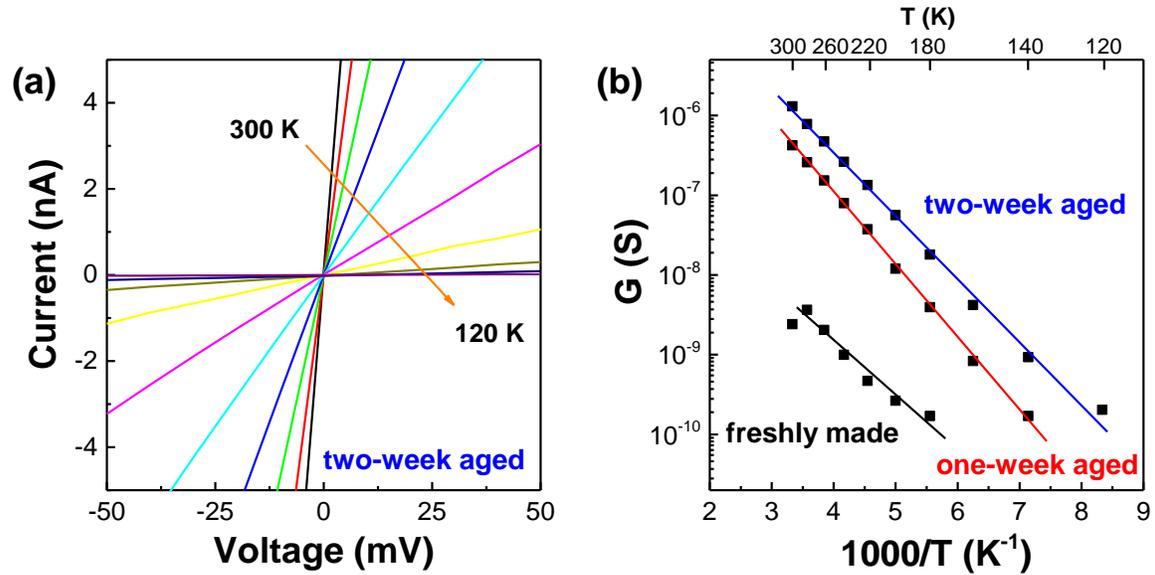

Figure 4. (a) Current-voltage (*I-V*) curves in the low bias regime (bias = 100 mV) at variable temperatures from 300 to 120 K for a typical film of 1-dodecene alkylated Si NCs after aging at room temperature for 2 week. Conductance decreases as temperature decreases. (b) Temperature dependence of the ohmic conductance when the film was freshly made, 1-week-aged and 2-week-aged in the range $120 < T < 300$ K. The data are displayed in log-linear scale and the error bar for the measurement of each data point is smaller than the symbol size used in this figure. Solid lines are linear fits for each aging stage.

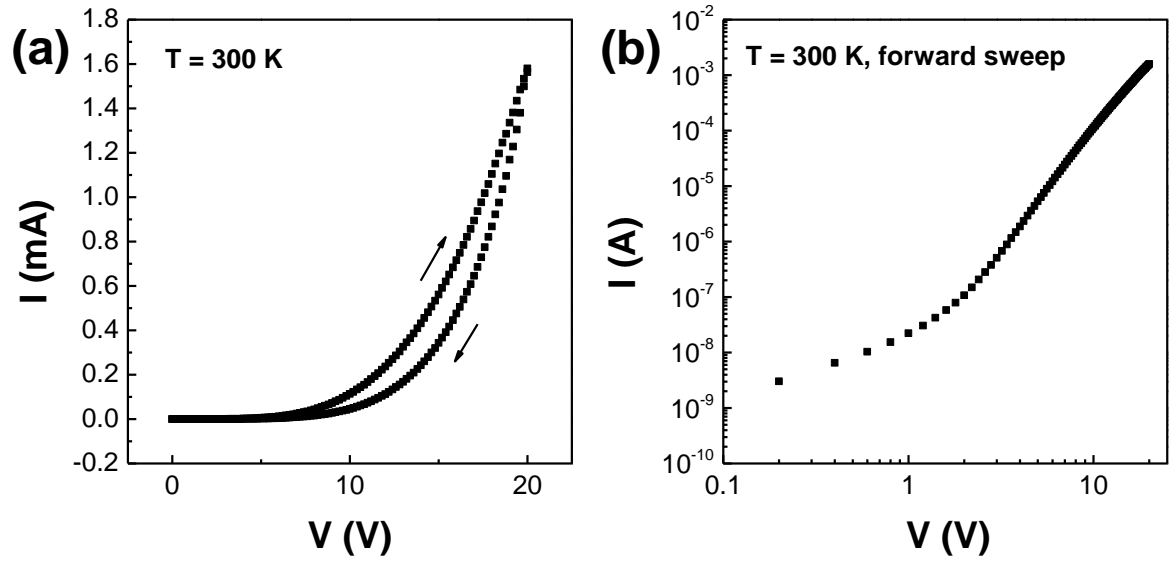

Figure 5. (a) Current-voltage (*I-V*) curves in the high bias regime at 300 K for a typical Si NC thin film after aging at room temperature for 2 week. Hysteresis can be seen from the forward and backward sweeps. (b) Forward sweep for *I-V* at 300 K in log-log scale.

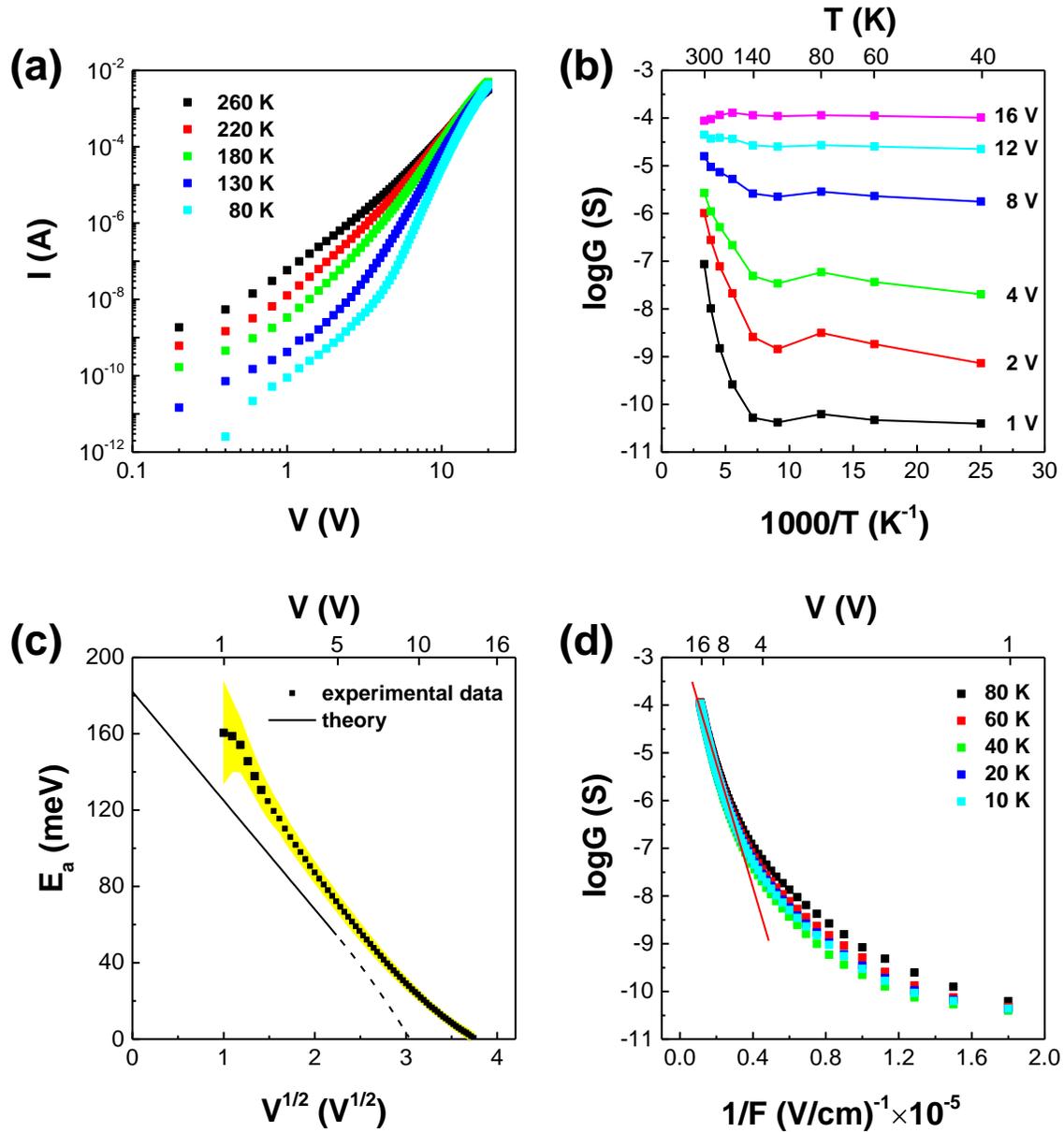

Figure 6. (a) Current-voltage (*I-V*) curves in the high bias regime at variable temperatures from 260 to 80 K for a typical Si NC thin film after aging at room temperature for 2 week. (b) Arrhenius plot of the temperature dependence of the film conductance at different voltages from 1 to 16 V. With voltage increasing, the slope of curves in the range of 300 to 140 K decreases. (c) Activation energy $E_a$ for the temperature range $140 < T < 300$ K at various voltages from 1 to 16 V. The black square is the activation energy fit from ln *G*

versus 1/*T* at each voltage point, and the yellow shadow represents the uncertainty of the activation energy caused by the linear fit. The solid black line shows the theoretical result up to 5 V described in eq 5 and the dash line indicates the linear dependence on voltage above 5 V. (d) Semi-log plot of conductance versus inverse electric field (1/*F*) for the same device in the temperature range 10 < *T* < 80 K and at bias voltages below 16 V. The red solid line is the linear fit for data from 5 to 16 V.

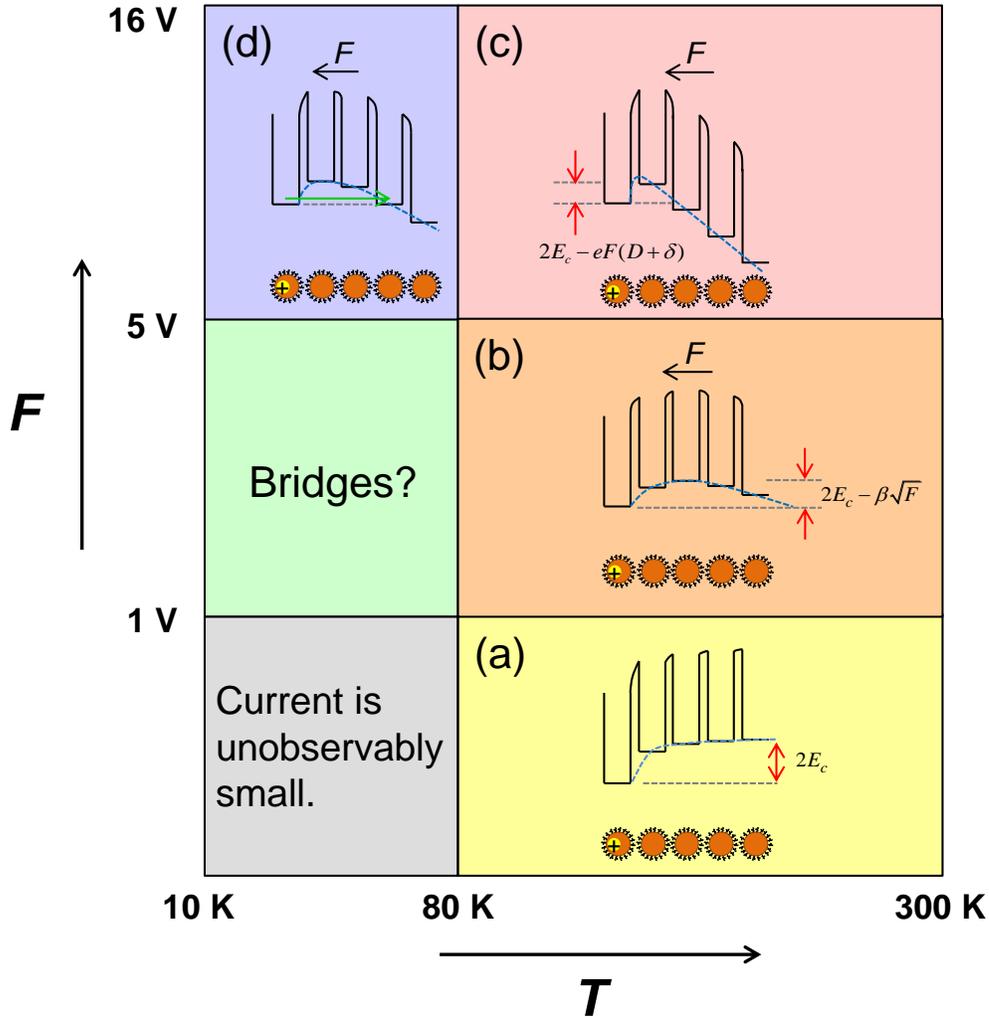

Figure 7. Ionization of a rare donor-containing NC. (a) $T > 80$ K and very weak electric field $F$ ($V < 1$ V). The ionization energy $2E_c$, necessary to create two charged NCs is shown. (b) $T > 80$ K and strong electric field $F$ ($1$ V $< V < 5$ V), which facilitates ionization, reducing ionization energy $2E_c$ by Poole-Frenkel mechanism ($\beta = \sqrt{e^3/4\pi\varepsilon_0\varepsilon_r}$). (c) $T > 80$ K and strong electric field $F$ ($5$ V $< V < 16$ V), the ionization energy linearly depends on the electric field. (d) $T < 80$ K and strong electric field $F$ ($5$ V $< V < 16$ V) when ionization happens by tunneling directly between non-neighboring NCs (cold ionization).